\documentclass{article}

\usepackage{amsmath, amssymb, amsthm}  

\usepackage{listings}  
\usepackage{geometry}  
\usepackage[utf8]{inputenc} 
\usepackage[T1]{fontenc}    
\usepackage{hyperref}       
\usepackage{url}            
\usepackage{booktabs}       
\usepackage{amsfonts}       
\usepackage{nicefrac}       
\usepackage{microtype}      
\usepackage{graphicx}
\usepackage{amsmath}
\usepackage{soul}
\usepackage[numbers,sort&compress]{natbib}  
\usepackage[affil-it]{authblk}\usepackage{boxedminipage}
\usepackage{fancyhdr}
\usepackage{lipsum} 

\title{\textbf{Intersubjective Model of AI-mediated Communication: Augmenting Human-Human Text Chat through LLM-based Adaptive Agent Pair}}

\author[a]{\textbf{Shutaro Aoyama}\footnote{These authors contributed equally to this work.}\footnote{sa4168@columbia.edu}}
\author[b]{\textbf{Rintaro Chujo}\textsuperscript{*}\footnote{chujo@nae-lab.org}}
\author[b]{\textbf{Ari Hautasaari}}
\author[b]{\textbf{Takeshi Naemura}}

\affil[a]{Columbia University}
\affil[b]{The University of Tokyo}

\date{}

\begin{document}
\maketitle

\vspace{-15pt}
\begin{abstract}
    The growing prevalence of Large Language Models (LLMs) is reshaping online text-based communication; a transformation that is extensively studied as AI-mediated communication. However, much of the existing research remains bound by traditional communication models, where messages are created and transmitted directly between humans despite LLMs being able to play a more active role in transforming messages. In this work, we propose the Intersubjective Model of AI-mediated Communication, an alternative communication model that leverages LLM-based adaptive agents to augment human-human communication. Unlike traditional communication models that focus on the accurate transmission of information, the Intersubjective Model allows for communication to be designed in an adaptive and customizable way to create alternative interactions by dynamically shaping messages in real time and facilitating shared understanding between the human participants. In this paper, we have developed a prototype text chat system based on the Intersubjective Model to describe the potential of this model, as well as the design space it affords.
\end{abstract}

\textbf{Keywords}: Computer Mediated Communication $|$ AI-Mediated Communication $|$ Large Language Models
\vspace{0.5cm}

\begin{figure}
    \centering
    \includegraphics[width=\textwidth]{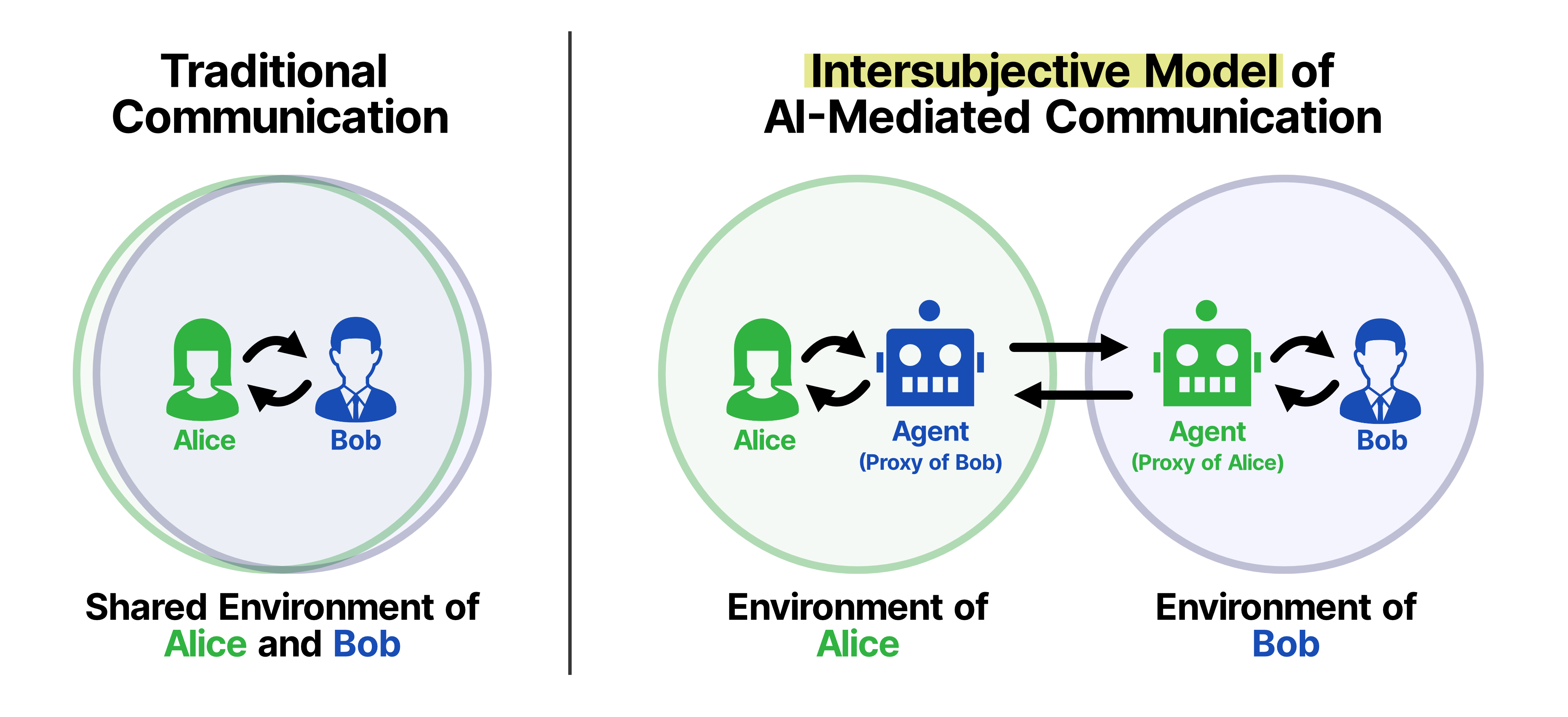}
    \caption{Conceptual diagram of communication under traditional model (left) and our Intersubjective Model (right).}
    \label{fig:teaser}
\end{figure}

\section{Introduction}
Models and tools for understanding and facilitating human communication have been developed across various fields for many years. Theories and models proposed in linguistics, psychology, sociology, and computer science have elucidated the essence of human communication and provided the foundation for achieving more effective dialogue. Technological advancements have fueled research particularly on computer-mediated communication (CMC), revealing the characteristics of communication through diverse digital platforms such as email, instant messaging, online forums, and social media. CMC transcends the limitations of time and space, significantly enhancing the efficiency and quality of information sharing and collaborative work.

However, these existing communication models and tools primarily assume that multiple individuals share the same objective environment and engage in direct communication. Even in the CMC field, research and implementation have focused on mechanisms where individuals communicate within shared environments such as chat rooms, video conferencing rooms, or threaded discussions. Conversely, developing communication models that remove the constraint of sharing an objective environment has remained largely unexplored.

In recent years, with the development of large language models (LLMs), a new area of AI-mediated communication (AIMC) has been attracting attention~\cite{Hancock2020}, and LLM-powered AIMC holds the potential to reshape communication paradigms. However, current AIMC research and implementations primarily focus on areas with limited AI autonomy, relying on human supervision for message suggestion and modification~\cite{Fu2024FromText, Mieczkowski2021AIMC, Hohenstein2023AIMC}. As such, these approaches can still be considered as supporting humans within the framework of "sharing an objective environment". 

This study proposes the \textit{Intersubjective Model} as an alternative AIMC framework. In the Intersubjective Model, individuals do not share a common environment; instead, they engage in conversations with AI agents within independent chat environments, where communication between humans occurs indirectly through information sharing between these AI agents. This model eliminates the constraint of sharing an objective environment, enabling a high degree of freedom to adaptively modulate message content and chronemics for each user. By designing agent-to-agent information sharing mechanisms and agent conversational capabilities, communication can be transformed to align with specific goals in the independent environments.

This research aims to elucidate the potential and challenges of AIMC based on the Intersubjective Model by proposing this novel approach and developing a prototype system. In the following sections, we first review related research on CMC and AIMC, highlighting areas that have not been yet sufficiently explored. We will then elaborate on the proposed Intersubjective Model, and discuss information transmission and feedback loops within the model as well as the design space it allows by drawing connections with existing communication theories. We will then detail the implementation of a prototype system based on the Intersubjective Model, as well as future research directions for AIMC. 


\section{Related Work}

\subsection{Computer-Mediated Communication}
With the proliferation of computers, computer-mediated communication (CMC) has become ubiquitous. Video conferencing systems like Zoom~\footnote{\url{https://zoom.us/}} and text chat tools such as Messenger~\footnote{\url{https://www.messenger.com/}} and WeChat~\footnote{\url{https://www.wechat.com/}} are widely used in both personal and professional settings. Consequently, the characteristics of CMC have been extensively studied in previous research.

One of the most prominent theories related to CMC is the Media Richness Theory~\cite{MediaRichness1986}. This theory posits that different media used for communication convey varying amounts of information, impacting the effectiveness of communication. For example, video conferencing transmits messages through diverse verbal and nonverbal cues such as facial expressions, gestures, and vocal tone, providing rich information from the speaker and real-time feedback from the listener. Conversely, text chats lack most of the nonverbal cue available in richer mediums, which can potentially hinder the communicative process leading to more messages having to be shared to achieve similar relational outcomes~\cite{WALTHER1992h,WALTHER1994a}. Furthermore, the level of media richness influences impression formation between interlocutors. The Hyperpersonal Model~\cite{HyperPersonal1996} suggests in the absence of rich verbal and nonverbal cues message recipients tend to project idealized impressions onto the message sender. The scarcity of information about the message sender and the context of the message can also lead to misunderstandings regarding the intended emotional tone of the message~\cite{Byron2008-km}. Thus, media choice in CMC is not merely a means of information transmission but a crucial factor that significantly affects interpersonal impression formation.

\subsection{AI-Mediated Communication}

Against the backdrop of CMC systems' pervasiveness in daily life, the emerging field of AI-Mediated Communication (AIMC) has gained attention as an extension of CMC.Hancock et al.~\cite{Hancock2020} define AIMC as "mediated communication between people in which a computational agent operates on behalf of a communicator by modifying, augmenting, or generating messages to accomplish communication or interpersonal goals". They emphasize AIMC's rapid expansion, and advocate for its examination across various dimensions, including the impact of AIMC on language, interpersonal dynamics, self-presentation, impression formation, trust, feedback, as well as long-term relationship formation and maintenance. Hancock et al. propose five key dimensions to describe AIMC systems: magnitude, autonomy, media type, optimization goal, and role orientation. In this research, we focus particularly on two of these dimensions: \textit{magnitude}, which reflects "the extent of the changes that AI enacts on messages", and \textit{autonomy}—"the degree to which AI can operate on messages without the sender's supervision".

The magnitude and autonomy dimensions provide a framework for understanding various AIMC applications. While magnitude primarily pertains to the technical capacity of the AI system to alter messages, autonomy refers to the interaction design, specifically the level of control that users have over the transformations applied by AI. These two dimensions create a matrix of possible AIMC systems with distinct characteristics. (Table \ref{table:aimc_matrix})

\begin{itemize}
    
    \item \textbf{Low Magnitude, Low Autonomy}: Examples of this combination include tools such as Grammarly~\cite{Grammarly}, which make small, incremental changes to messages and allow users to retain control over the content. Previous research, such as that by Fu et al.~\cite{Fu2024FromText}, has explored the short-term and long-term effects of AI-powered writing assistance. Studies by Jakesch et al.~\cite{Jakesch2019AIMC}, Hohenstein et al.~\cite{Hohenstein2023AIMC}, and Mieczkowski et al.~\cite{Mieczkowski2021AIMC} have shown that these types of AIMC systems can introduce subtle biases and impact perceptions of trustworthiness.

    \item \textbf{High Magnitude, Low Autonomy}: Here, AIMC systems enact substantial transformations but maintain user oversight. This configuration includes technologies for complete voice or facial reconstruction and applications that entirely rewrite messages or transform input content into a different format. Examples include tools such as Gmail's Smart Reply~\cite{GmailSmartReply}, which can significantly alter message phrasing while allowing users to retain control. Algouzi et al.~\cite{Algouzi2023Gmail} investigated the sociocultural implications of using such tools. Harashima's work on intellectual coding~\cite{harashima1991intelligent} has also explored significant transformation based on encoding and decoding facial expressions and voice audios. While these systems can perform large modifications, they rarely employ high autonomy, and users usually retain control over significant alterations.
    
    \item \textbf{Low Magnitude, High Autonomy}: In this quadrant, the AI applies relatively minor modifications autonomously, without requiring user supervision for each transformation. For example, Arias-Sarah et al. ~\cite{sarah2024face}  modified the facial expressions in video conferences without notifying the participants. Certain creative applications have used simple algorithms to rephrase or adjust chat messages dynamically (e.g.,~\cite{mwitm}). Systems with this configuration are characterized by a limited degree of message transformation but function independently of constant user input, thus requiring higher autonomy than the previous category. Moreover, AI Clone Agents, as depicted in the science fiction series "Black Mirror~\footnote{\url{https://www.imdb.com/title/tt20247352/}}" and currently under development by various startups (e.g., Synthesia~\footnote{\url{https://www.synthesia.io/}}), can also be considered to be situated within this quadrant.
    
    \item \textbf{High Magnitude, High Autonomy}: In this category, the system autonomously performs extensive modifications on messages, often without user involvement in the process. In this case, each user experiences a significant transformation to their messages, and the notion of a single objective world shared by all participants is replaced by distinct subjective environments mediated by AI. As such, this configuration offers considerable potential for AI to mediate communication. However, research in this area remains limited. Our study proposes the \textbf{Intersubjective Model} as a framework for this quadrant, exploring the effects of high magnitude, high autonomy transformations within the context of text-based communication.
\end{itemize}
\begin{table}[h]
\centering
\begin{tabular}{c|c|c}
\textbf{} & \textbf{Low Magnitude} & \textbf{High Magnitude} \\ \hline
\textbf{Low Autonomy} 
& \begin{tabular}[c]{@{}c@{}}Grammarly~\cite{Grammarly} \\ Fu et al.~\cite{Fu2024FromText} \\ Jakesch et al.~\cite{Jakesch2019AIMC} \\ Hohenstein et al.~\cite{Hohenstein2023AIMC} \\ Mieczkowski et al.~\cite{Mieczkowski2021AIMC} \end{tabular} 
& \begin{tabular}[c]{@{}c@{}}Gmail's Smart Reply~\cite{GmailSmartReply} \\ Algouzi et al.~\cite{Algouzi2023Gmail} \\ Harashima~\cite{harashima1991intelligent} \end{tabular} \\ \hline

\textbf{High Autonomy} 
& \begin{tabular}[c]{@{}c@{}}Arias-Sarah et al.~\cite{sarah2024face} \\ MWITM~\cite{mwitm} \end{tabular} 
& \begin{tabular}[c]{@{}c@{}}Intersubjective Model \\ \end{tabular} \\ 
\end{tabular}
\caption{AIMC Systems Categorized by Autonomy and Magnitude}
\label{table:aimc_matrix}
\end{table}




\section{The Intersubjective Model}

This study proposes the \textit{Intersubjective Model} as a novel communication model for AIMC. Conventional AIMC assumes that two individuals share the same environment and engage in direct communication. In contrast, our proposed model differs significantly in that each participant exists in an independent environment and communicates through LLM-based adaptive agent for the other party.

\begin{figure}
  \includegraphics[width=\textwidth]{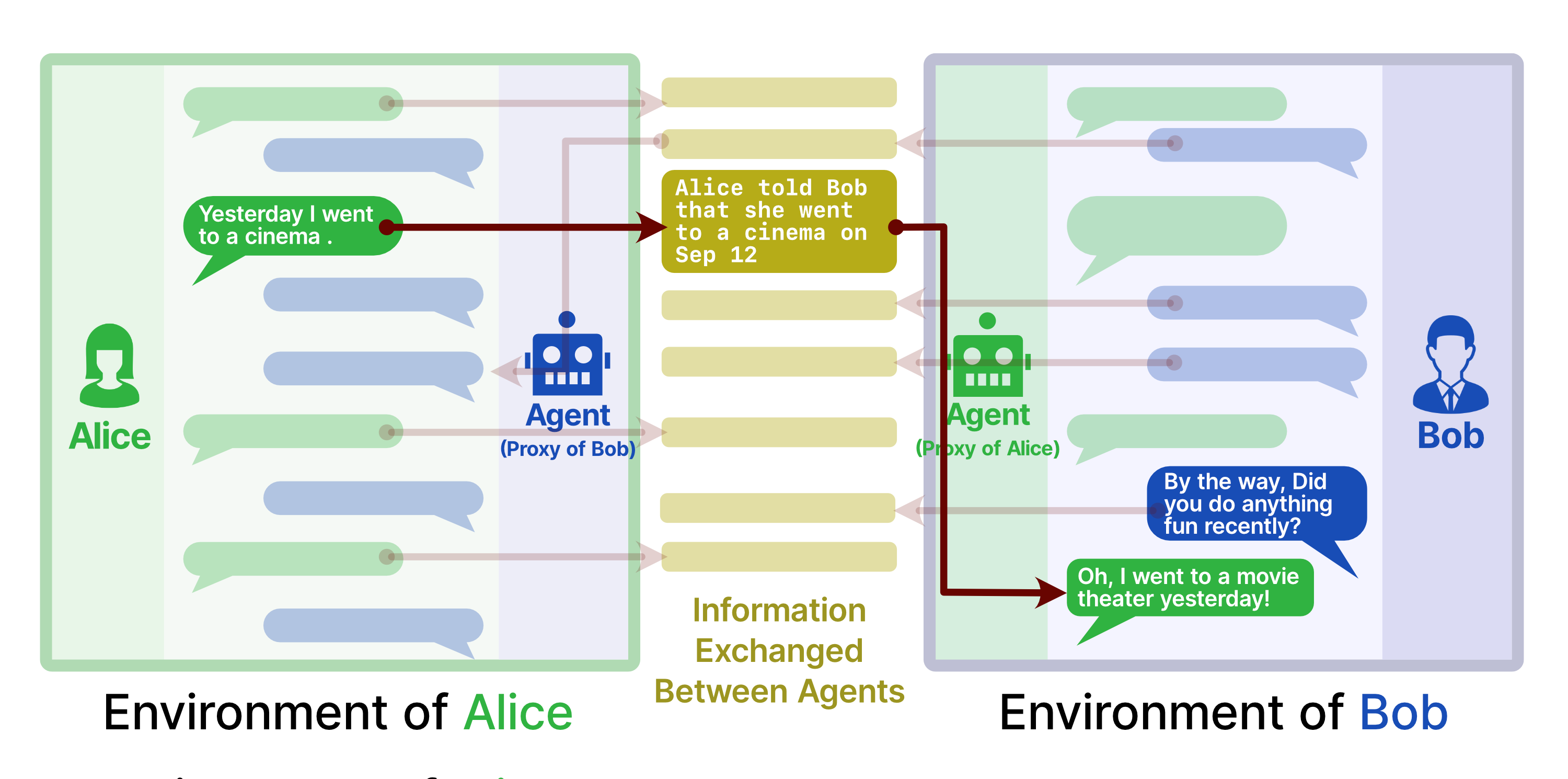}
  \caption{Process of a message sent by Alice transmitted to Bob, mediated by agents. (1) Alice says, "Yesterday I went to a cinema." (2) Agent A in Alice's environment (i.e., proxy of Bob) extracts the information "Alice told Bob that she went to a cinema on Sep. 12" and transmits it to Agent B in Bob's environment (i.e., proxy of Alice). (3) Agent B adds this shared information to its knowledge base. (4) Later, Agent B says, "Oh, I went to a movie theater yesterday!" and Bob receives this message.}
  \label{fig:transmission}
\end{figure}

The dyadic communication process in the Intersubjective Model consists of two interactions:
\begin{itemize}
\item \textbf{Dialogue between human and their agent in each individual's environment (Figure \ref{fig:transmission}, left and right):} Each environment contains a human and an agent. The human interacts with this agent, which acts as an agent for the other person. This structure allows individuals to communicate indirectly with another human while each individual occupies a completely different environment.
\item \textbf{Information transmission between agents (Figure \ref{fig:transmission}, center):} Information is transferred between agents in different environments. Each agent extracts information from its conversation with the human and transmits it to the agent in the other environment. This enables each agent to continue the conversation with its human while possessing information from the other environment.
\end{itemize}

In the Intersubjective Model, a message transmission occurs as follows (assuming Alice is the sender, Bob is the receiver, Agent A is a proxy of Bob in Alice's environment, and Agent B is a proxy of Alice in Bob's environment):
\begin{enumerate}
\item Alice and Agent A are engaged in an ongoing conversation.
\item Agent A extracts a piece of information from the conversation with Alice.
\item Agent A transmits the piece of information to Agent B.
\item Agent B integrates the received information into its knowledge base.
\item Bob and Agent B are engaged in an ongoing conversation, and Agent B brings up the information extracted by Agent A from Alice at an appropriate time.

\end{enumerate}

This process occurs bidirectionally and continuously. As a result, participants can partially share each other's subjective environments through their agents while maintaining their own subjective spaces. This enables them to engage in what can be described as intersubjective communication.

\subsection{Possible Use Cases of the Intersubjective Model}
The benefit of this model is its ability to support the achievement of communication goals by designing agents tailored to those goals. 
Agents can intervene in the communicative process through two functions:

\begin{itemize}
\item \textbf{Extraction:} Extracting information from communication and sharing it with other agents.
\item \textbf{Conversation:} Communicating with the human while incorporating information shared by other agents.
\end{itemize}

In Extraction, the criteria for determining what to extract and what to omit can be designed according to the objective. For instance, agents can be designed to extract the maximum meaning and nuances of a message or to selectively extract specific information while discarding the rest.

In Conversation, agents can be designed to reflect and act upon the information extracted and shared from the other environment in a manner aligned with the communication goal. This involves adjusting elements such as content, style, and timing of message delivery. For example, an agent could be designed to echo Alice's message immediately in Bob's environment or to hold back the information and introduce it later in a contextually appropriate manner. Agents can also be designed to behave differently towards Alice and Bob.

\begin{figure}[ht]
\centering
\includegraphics[width=.48\textwidth]{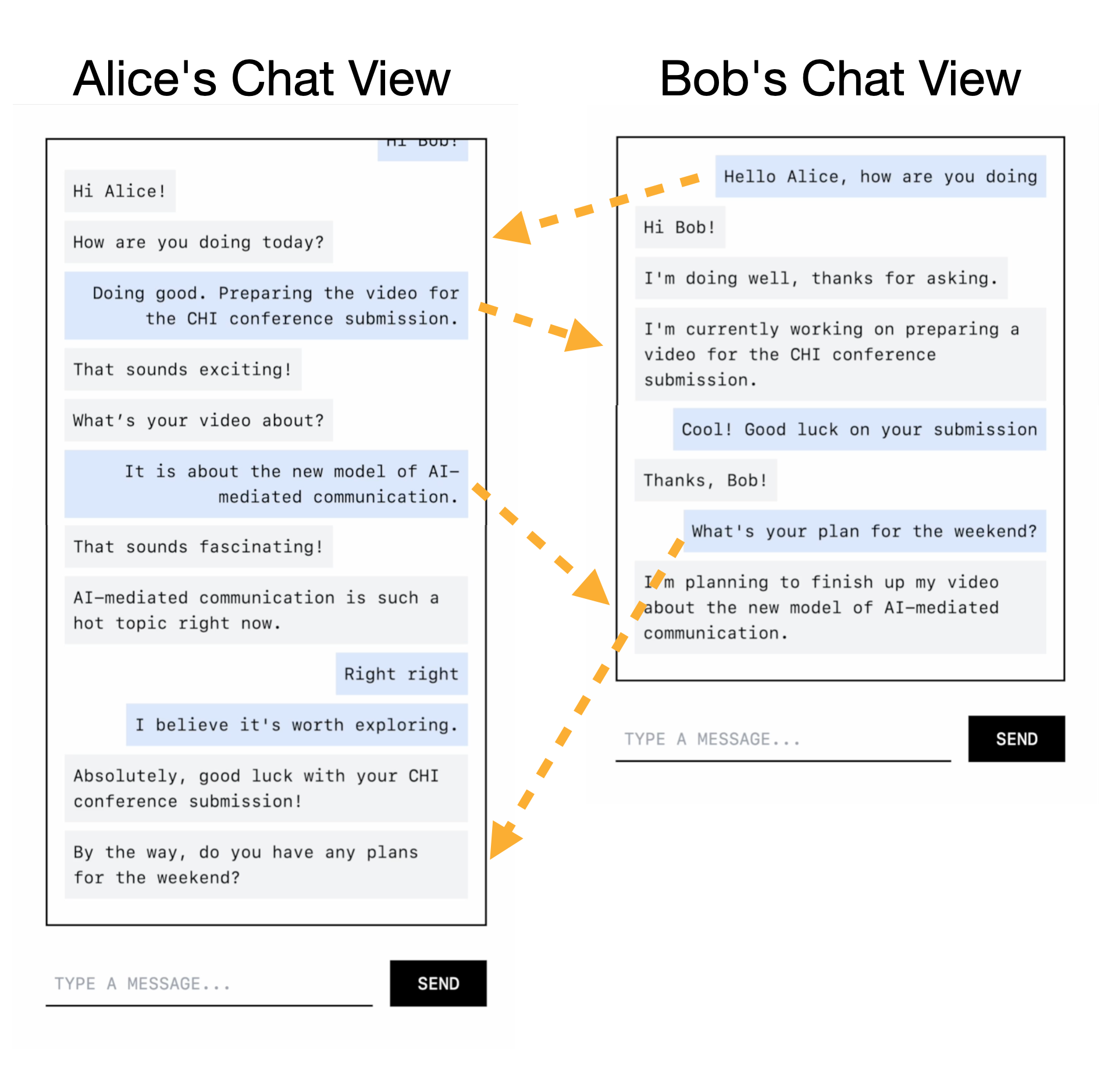}
  \caption{Example 1 - Lively Conversation}
  \label{fig:ex1}
\end{figure}

Below are examples of goal settings and corresponding agent designs:

\subsubsection{Example 1: Lively Conversation (Fig.~\ref{fig:ex1})}
\begin{itemize}
\item \textbf{Goal:} To foster a lively and engaging conversation.
\item \textbf{Extraction:} Omit information about emotional intensity and extract only the content of the utterances.
\item \textbf{Conversation:} Express the transmitted content with consistently high enthusiasm. Overreact to the human's utterances. Respond quickly.
\end{itemize}

\subsubsection{Example 2: Focused Discussion}
\begin{itemize}
\item \textbf{Goal:} To facilitate a focused and productive discussion on a specific topic.
\item \textbf{Extraction:} Extract only information relevant to the topic.
\item \textbf{Conversation:} Express the transmitted content by relating it to the topic. Respond to the human's utterances by connecting them to the topic. Intentionally slow down the response speed.
\end{itemize}
Agents can also be designed to behave differently towards each individual to meet personalized communication needs. Here are some examples:

\subsubsection{Example 3: Overcoming Language Barriers}
\begin{itemize}
\item \textbf{Goal:} To facilitate communication across language barriers.
\item \textbf{Extraction:} Extract messages while distilling meaning from language-specific expressions.
\item \textbf{Conversation:} Translate the extracted messages into each individual's native language.
\end{itemize}

\begin{figure}[ht]
\centering
\includegraphics[width=.65\textwidth]{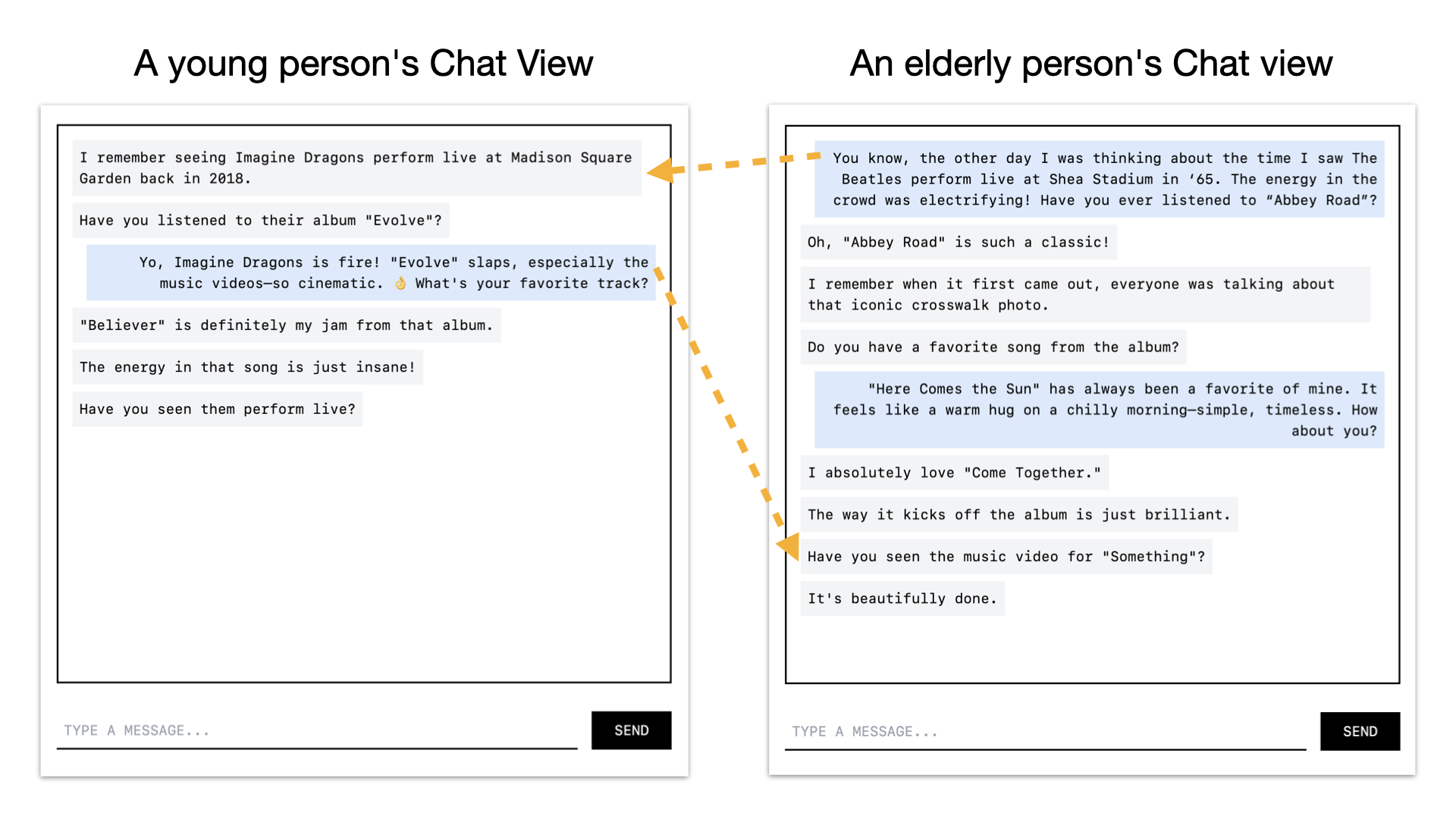}
  \caption{Example 4 - Bridging Generational Gaps}
  \label{fig:ex4}
\end{figure}

\subsubsection{Example 4: Bridging Generational Gaps (Fig.~\ref{fig:ex4})}
\begin{itemize}
\item \textbf{Goal:} Bridge the generational gap in conversations between teenagers and octogenarians.
\item \textbf{Extraction:} Extract messages while distilling meaning from generation-specific expressions.
\item \textbf{Conversation:} Communicate the extracted messages using expressions that resonate with each generation.
\end{itemize}

\begin{figure}[ht]
\centering
\includegraphics[width=.65\textwidth]{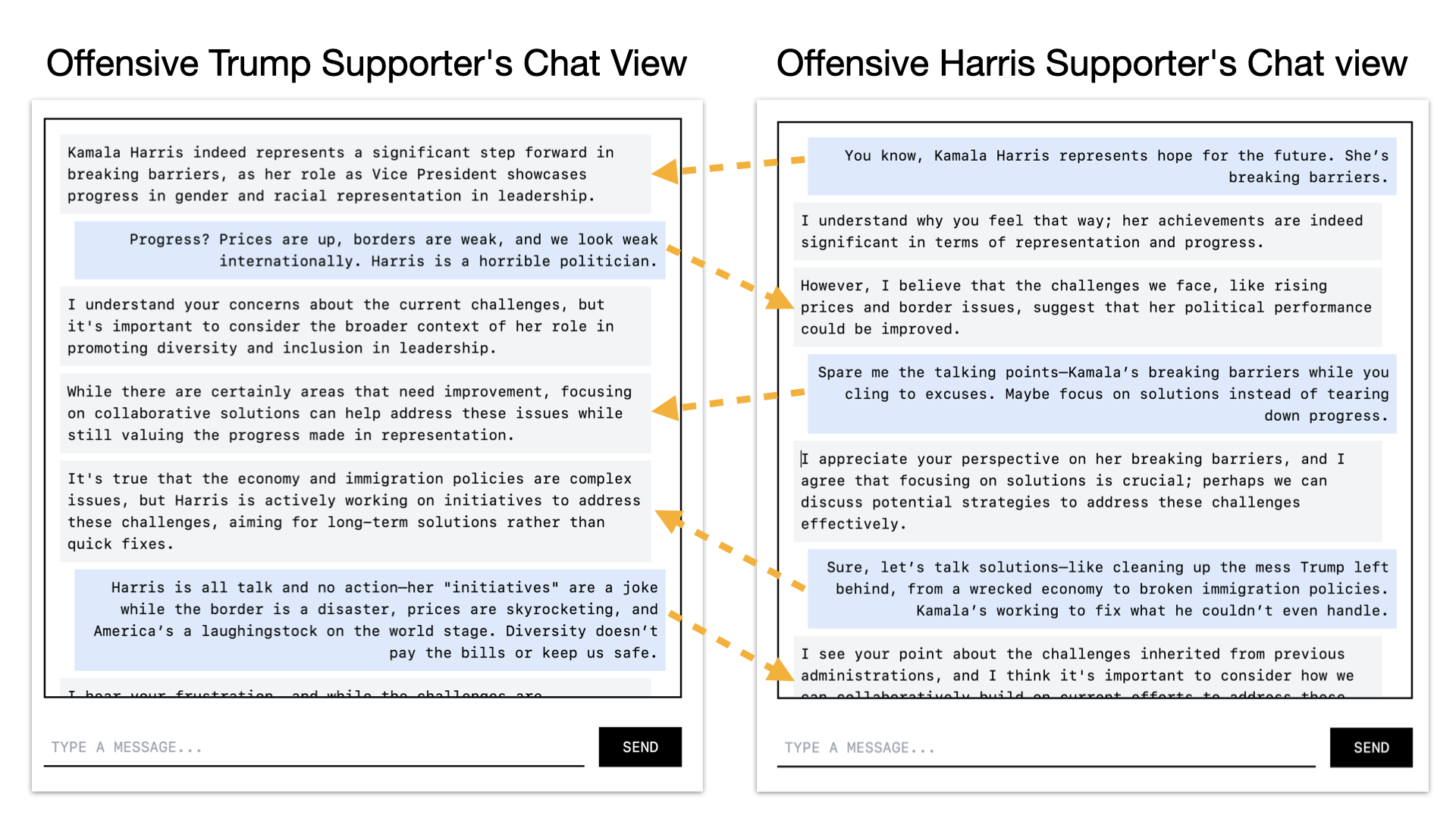}
  \caption{Example 5 - Facilitating Calm Discussion}
  \label{fig:ex5}
\end{figure}

\subsubsection{Example 5: Facilitating Calm Discussion (Fig.~\ref{fig:ex5})}
\begin{itemize}
\item \textbf{Goal:} To facilitate a calm discussion between two individuals with differing political views who tend to engage in emotional arguments.
\item \textbf{Extraction:} Omit information about emotional outbursts and attitudes, focusing only on extracting each humans' core arguments.
\item \textbf{Conversation:} Communicate the extracted arguments in an empathetic and understanding manner.
\end{itemize}
As demonstrated, by designing the extraction mechanisms and agent behaviors according to specific goals, the Intersubjective Model can support the achievement of diverse communication goals.

\subsection{Theoretical Understanding}
This chapter examines how traditional models explain objective-based and goal-oriented communication and applies those explanations to the Intersubjective Model.

\subsubsection{Transmission of Messages in Traditional Models}


\begin{figure}
  \includegraphics[width=\textwidth]{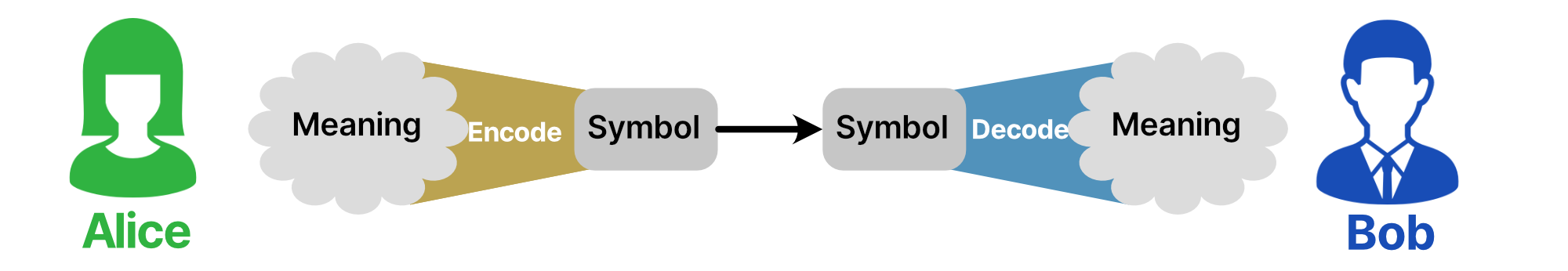}
  \caption{Transission process in the Schramm's model of communication. }
  \label{fig:schramm}
\end{figure}

Traditional communication models often describe information transmission as a process where humans encode and decode meaning into symbols. Models like those proposed by Shannon~\cite{Shannon1948}, Weaver~\cite{Weaver1953}, and Schramm~\cite{Schramm1954} exemplify this approach.

Shannon~\cite{Shannon1948} proposed a communication model that forms the basis of information theory. Weaver~\cite{Weaver1953} extended Shannon's model to human communication and categorized the encoding/decoding process into three levels:

\begin{itemize}
\item Level A: How accurately can symbols be transmitted in the process of a machine encoding, transmitting, and decoding them?
\item Level B: How accurately can meaning be conveyed when the sender encodes meaning into symbols and the receiver decodes symbols into meaning?
\item Level C: How effectively can intentions be communicated to produce the desired effect when humans encode their intended effects into meaning and the receiver decodes meaning into effects?
\end{itemize}
Schramm~\cite{Schramm1954} presented a similar communication model, explaining how the meaning of an idea is encoded into symbols by the sender, transmitted, and then decoded back into an idea by the receiver. (Figure \ref{fig:schramm}) He also highlighted the potential for errors in this process. Errors arise when information omitted during encoding is not accurately supplemented during decoding, particularly when there is a lack of a shared field of experience, such as language, experience, or values. For example, an American who has never studied Russian cannot encode or decode messages in Russian. Similarly, an indigenous person who has never heard of airplanes cannot accurately decode a message about airplanes.

In summary, traditional models explain information transmission in communication as a process of encoding and decoding meaning into symbols.

\subsubsection{Transmission of Messages in the Intersubjective Model}

\begin{figure}
  \includegraphics[width=\textwidth]{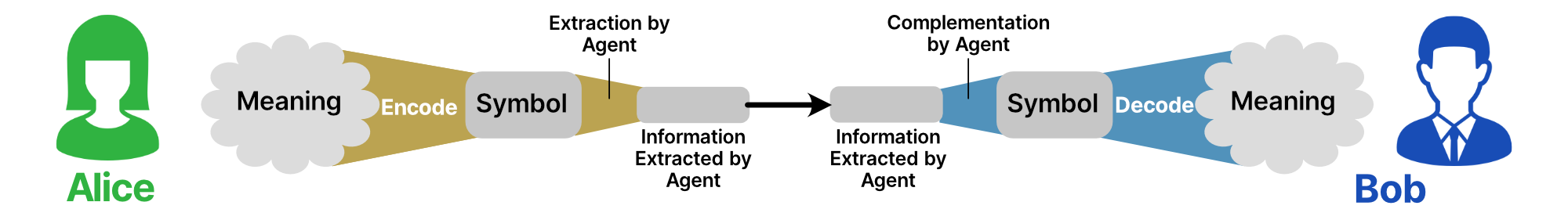}
  \caption{Transmission process in the Intersubjective Model of communication. }
  \label{fig:schramm_ours}
\end{figure}

Building upon the explanations from traditional models, we now describe information transmission within our proposed model.

While in traditional models only humans perform the "meaning-handling encoding/decoding," in the Intersubjective Model, LLMs also participate in this process. In encoding, LLMs extract information from existing text (already symbolized) and further encode it. In decoding, LLMs supplement missing information to generate utterances. This introduces computer intervention into the process of meaning omission and supplementation, which was previously handled solely by humans. (Figure \ref{fig:schramm_ours})

The Intersubjective Model allows for the intentional design of errors that occur during the meaning-symbol encoding/decoding process. Unlike the traditional paradigm that assumes accurate symbol transmission, our model enables purposeful omission and supplementation of meaning during symbol transmission.

This differs from the encoding/decoding errors in the traditional paradigm, which aimed for "correct transmission". The Intersubjective Model performs active and purposeful adjustments to information based on the communication goals. Therefore, instead of "errors", we refer to the changes in information caused by encoding/decoding as \textit{modulation}.

Modulation can be operationalized by abandoning correct transmission in multiple layers:
\begin{enumerate}
\item \textbf{Abandoning "correct transmission of symbols":} When Schramm's shared experience is absent in the original communication, making encoding/decoding inherently difficult, modulation can bridge the gap. For example, Harashima's concept of intellectual coding~\cite{harashima1991intelligent} aligns with this idea: "If conceptual information and semantic information are extracted from linguistic information, it would be possible to realize automatic translation communication between different languages based on that conceptual and semantic information." This suggests that modulation at the symbol level can be employed to ensure accurate transmission of information at higher levels (meaning, concepts).


\item \textbf{Abandoning "correct transmission of meaning":} In some cases, introducing modulation itself aligns with the communication goal. Examples include enhancing conversation liveliness, facilitating focused discussions, and mediating political debates. Selective self-presentation in the Hyperpersonal Model\cite{HyperPersonal1996}, where omission and biased supplementation of information due to computer mediation lead to positive impressions, is a related phenomenon. The Intersubjective Model allows for the intentional design of such effects.

\end{enumerate}
\subsubsection{Feedback Loop in Traditional Models}

\begin{figure}
  \includegraphics[width=\textwidth]{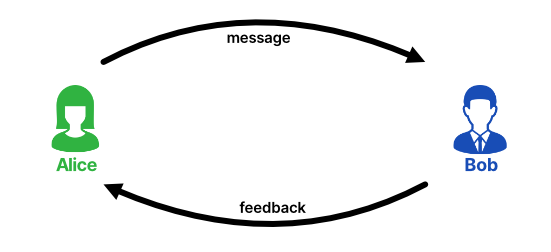}
  \caption{Feedback loop in the Schramm's Model of communication.}
  \label{fig:feedback}
\end{figure}

Traditional communication models offer various explanations for the overall picture of interaction. In contrast to the unidirectional communication models of Shannon and Weaver, several models propose an understanding based on bidirectional loops~\cite{barker1966model, Schramm1954}. For instance, Schramm's communication model~\cite{Schramm1954} views interaction between two individuals as a feedback loop. (Figure \ref{fig:feedback})

In Schramm's model, feedback on a message supports accurate information transmission. The sender can modify their message based on the receiver's response, including verbal replies and nonverbal cues such as facial expressions. This feedback loop mitigates the impact of noise during transmission and errors during encoding/decoding, leading to more accurate information transfer.

\subsubsection{Feedback Loop in the Intersubjective Model}

\begin{figure}
  \includegraphics[width=\textwidth]{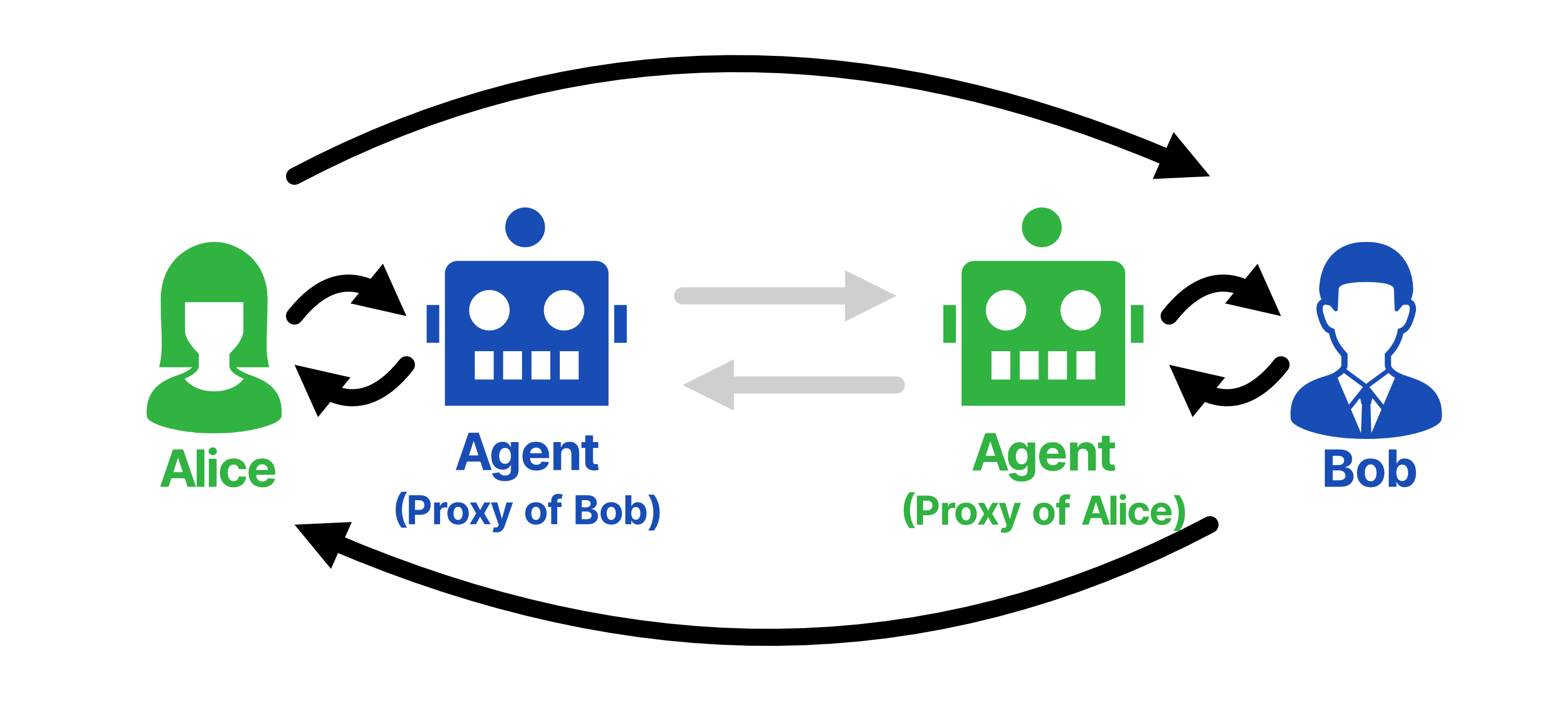}
  \caption{Feedback loop in the Intersubjective Model of communication. }
  \label{fig:feedback_ours}
\end{figure}

Building on traditional feedback loop concepts, the Intersubjective Model reshapes how feedback dynamics operate by leveraging AI agents to mediate interactions. As shown in Figure \ref{fig:feedback_ours}, we can consider the communication between Alice and Bob as involving two distinct feedback loops: 

\begin{enumerate}
    \item The \textbf{Inner Feedback Loop} between Alice and her Agent (proxy of Bob) and between Bob and his Agent (proxy of Alice).
    \item The \textbf{Outer Feedback Loop} between Alice and Bob.
\end{enumerate}

In the Inner Loop, human participants receive immediate, context-sensitive feedback through responses generated by their respective agents. These exchanges often involve rapid acknowledgment cues or brief replies that keep the conversation moving. Since the agents can respond independently of the human interlocutors, this loop accelerates the communication rhythm, offering timely and relevant feedback based on the ongoing dialogue. For example, if Alice says, “I went hiking yesterday,” her Agent may instantly acknowledge with a contextual response like, “That sounds refreshing!” The speed of this feedback gives the interaction a sense of synchrony, even in asynchronous communication scenarios.

Conversely, the Outer Feedback Loop involves the transmission of more substantial information between agents, which represents the core meaning of the human participants' exchanges. This loop handles the extraction, encoding, and eventual delivery of intended meanings. In this loop, the conversation is modulated in accordance with the AI agents’ interpretation and the specific communication goals. This aspect of the feedback process ensures that larger conversational intent is preserved, even as AI handles the faster-paced responses.

By separating feedback into these two loops, the Intersubjective Model allows AI agents to provide instantaneous feedback on specific elements of the conversation while maintaining the fidelity of the message transmission at a broader scale. Traditional models rely on human-driven feedback loops for meaning confirmation and adjustment, but our model enables purpose-driven, high-speed feedback loops managed by AI, enhancing the flow of communication. Additionally, the AI-driven pacing can impart a sense of synchrony, even in asynchronous exchanges, thus creating a more cohesive and engaging conversational experience.

\section{Implementation}

\begin{figure}
  \includegraphics[width=\textwidth]{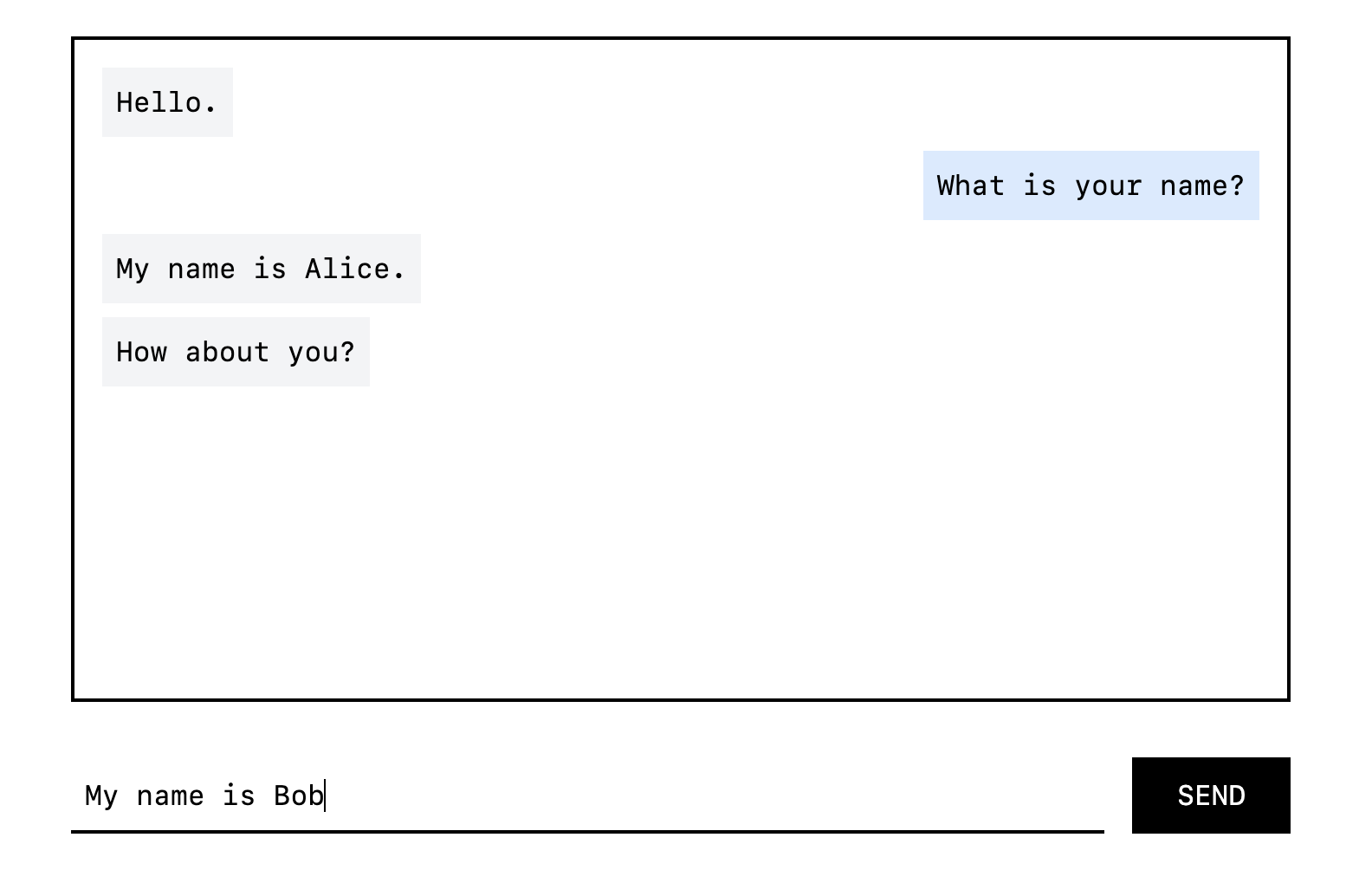}
  \caption{Online Chat Interface}
  \label{fig:chat}
\end{figure}

This section describes the design and implementation of a prototype text chat system 
based on the Intersubjective Model. The system was designed in Japanese, and the following sections include translations by the authors where applicable. 
The objective of this prototype is to partially demonstrate the effectiveness of the model, as well as to gain insights into the mental models, effects, and challenges that arise when users interact with and through the system.  The prototype system comprises two main components: an online chat interface for user interactions, and agents that mediate each user's environment.

\subsection{Online Chat Interface}

The online chat interface is designed with minimal functionality for simplicity (Figure \ref{fig:chat}). Users can view messages sent by other humans or agents, and input and send their own messages. The interface is built using React, and Supabase is used as the real-time database.

\subsection{Agents}

The agent in each user's environment has two core functions: Extraction and Conversation. Each agent interacts with its respective user while transmitting extracted information to the agent in the other user's environment. 

\subsubsection{Extraction}

The Extraction function extracts information from the conversation between the human and the agent and shares it with other agents. It takes two inputs: 1) the message log and 2) instructions on what information to extract. These inputs are fed into an LLM, which outputs the extracted information. This information is then shared with other agents. We used OpenAI's GPT-4o model as the LLM for this function.

\subsubsection{Conversation}

The Conversation function manages the conversation between the agent and the user. It monitors information sent by other agents and messages from the human, sending messages to the human at appropriate times.

Outgoing messages are generated by an LLM. The generation process takes three inputs: 1) information received from other agents, 2) the conversation log with the human, and 3) guidelines for generating responses aligned with the communication objective. We used OpenAI's GPT-4o model as the LLM for message generation.

\subsection{Enabling Intersubjective Communication}

Our prototype system is designed to enable a text chat communication environment based on the Intersubjective Model. The specific design and tuning of the Extraction criteria and conversation timing control, tailored to achieve specific communication objectives, are described in the following sections. 

\section{Example: Enhancing First-Time Conversations}

This section describes the design and implementation of a prototype text chat system aimed at enhancing the communicative experience between individuals meeting and interacting for the first time using a system tuned based on the Intersubjective Model. We designed the agent's Extraction and Conversation functions, including prompts and response timing control algorithms, to achieve the goal of "a better conversation experience for people meeting for the first time".

\subsection{Extraction}

The goal of Extraction in this context is simply to extract information from a message. Therefore, we employed a simple prompt instructing the LLM to describe the message.

\paragraph{System Prompt:}
\begin{quote}
    {\small
    \texttt{You are an AI assistant for conversation analysis.}}
\end{quote}
\paragraph{User Prompt:}
\begin{quote}
    {\small
    \texttt{See the latest message above and describe it.\\
\\
Example: "Tom is talking about how excited he is for his trip to Tokyo next week."\\
If there is no content or meaning, respond "NONE" without any extra comments.\\}}
\end{quote}

\subsubsection{Validation}

Similarly, the validation prompt is designed to simply check whether the agent has mentioned the described information.

\paragraph{System Prompt:}
\begin{quote}
    {\small
    \texttt{You are an AI assistant that analyzes conversations. Please analyze the given conversation log and determine whether a specific event has occurred.}}
\end{quote}
\paragraph{User Prompt:}
\begin{quote}
    {\small
    \texttt{Conversation Log:\\
\$\{chatContext\}\\
\\
Please determine if the following event has occurred in the above conversation.\\
If it seems to have occurred, answer "true"; if it has not, answer "false".\\
\\
Event: \$\{event.content\}\\
\\
Answer:}}
\end{quote}

\subsection{Conversation}

\paragraph{Response Generation Prompt:}

We tailored the response guidelines in the prompt to align with the goal of fostering natural and engaging conversations. Notably, we included the instruction "Do not start the message with expressions of agreement or empathy" to prevent the LLM from generating generic and potentially insincere expressions of empathy, which is a tendency for GPT-4o observed in our pilot study.

\paragraph{System Prompt:}
\begin{quote}
    {\small
    \texttt{Please generate *one* response as if you were \$\{username\}.\\
\\
*GOAL*\\
You are chatting with someone for the first time.\\
Make the partner feel, "This person could be a friend of mine," "I could have a pleasant conversation with this person".\\
\\
*RESPONSE GUIDELINES*\\
Respond actively but focus on building upon the partner's topic before introducing a new one.\\
Do not start the message with expressions of agreement or empathy.\\
When relevant, naturally introduce related information or questions to help deepen the conversation.\\
Avoid multiple expansions of the conversation in a single response; focus on one idea at a time.\\
Ensure responses feel natural, engaging, and human-like, without overloading the conversation.\\
Respond in the same language as the partner's message.\\
Keep the message short, maximum 3 sentences.\\
\\
*IMPORTANT RULES*\\
Respond in the same language as the partner's message.\\
Address some of the given tasks so that those will happen in the conversation.\\
When you receive a question and the task information is not available to correctly answer the question, give an evasive response.\\
If sending a new message as \$\{username\} is not needed, respond with "SKIP".\\}}
\end{quote}
\paragraph{User Prompt:}
\begin{quote}
    {\small
    \texttt{\# Tasks to complete\\
\$\{taskList\}\\
\\
\# Chat History\\
\$\{userChatHistory\}\\
\$\{username\}: \{NEW MESSAGE HERE\}\\
\\
\# Instruction\\
Generate one new message as if you were \$\{this.getUsernameFromUserId(user)\}.\\
\\
\# Format\\
\$\{username\}: \{NEW MESSAGE HERE\}}}
\end{quote}

\paragraph{Response Timing Control:}

We aimed to create a natural and engaging conversational flow by providing fast feedback while avoiding unnatural, robotic-like behavior.

When new information arrives, either from an AI agent or a human message, the system immediately clears any pending messages in the queue. It then generates a new response based on the current context, taking advantage of the fast generation speed of the underlying language model (GPT-4o) to provide near-instantaneous replies.

The generated response is split into individual sentences based on punctuation marks, and these sentences are added to a queue for sending. The main loop of the system continuously checks the queue, waiting for a certain duration before sending each message. The waiting time (s) is 2.5 times the number of characters in the message, assuming Japanese text. For example, a 10-character Japanese message would result in a 25-second delay.

This delay factor was determined through a pilot study to minimize perceived unnaturalness in the conversation. By sending messages sentence by sentence with predetermined delays, the system aims to mimic a more human-like pace of communication, avoiding a robotic or unnatural feel.

The overall goal is to strike a balance between providing fast, responsive feedback to keep the conversation engaging while also introducing pauses and delays that feel natural to maintain a comfortable, human-like conversational rhythm.

\section{Call for Future Work}
This section outlines potential avenues for future research building upon the findings and contributions of this study.

\subsection{Empirical Evaluation and User Studies}

To further validate and refine the Intersubjective Model, empirical research is crucial. Future studies should investigate:

\begin{itemize}
\item \textbf{Controlled experiments:} Conducting controlled experiments to evaluate the impact of different agent designs on communication outcomes, such as understanding, rapport, and task performance. This includes comparing the Intersubjective Model to traditional communication paradigms. Specific manipulations could include varying levels of agent autonomy, magnitude of message modification, and communication goals.
\item \textbf{Field studies:} Deploying prototype systems in real-world settings to understand how the Intersubjective Model functions in naturalistic communication contexts and to gather user feedback on its usability and effectiveness. These studies can focus on diverse user groups and communication scenarios, including conversations between strangers, collaborative work, and cross-cultural communication.
\item \textbf{Longitudinal studies:} Examining the long-term effects of using intersubjective communication systems on interpersonal relationships, communication patterns, and individual well-being.
\end{itemize}

\subsection{Ethical and Societal Implications}

The widespread adoption of intersubjective communication systems raises important ethical considerations. Future research should address:

\begin{itemize}
\item \textbf{Transparency and control:} Developing mechanisms to ensure transparency in agent behavior and provide users with control over how their messages are modulated.
\item \textbf{Bias and fairness:} Investigating potential biases in agent design and developing strategies to mitigate these biases and ensure fairness in communication.
\item \textbf{Privacy and security:} Addressing privacy and security concerns related to the collection and use of personal data by intersubjective communication systems.
\item \textbf{Social impact:} Analyzing the broader societal implications of widespread adoption of intersubjective communication, including its potential impact on social cohesion, political discourse, and human interaction.
\end{itemize}

\subsection{Expanding the Structure of the Intersubjective Model}
While the Intersubjective Model presented in this study focused on dyadic communication, its structure holds potential for further expansion. For instance, increasing the number of participants could enable the model to handle communication involving groups of 3-10 individuals, dozens to hundreds of individuals, or even larger scales of thousands to millions.

Moreover, non-bidirectional communication structures could be explored. For example, it might be possible to create scenarios where person A believes they are talking to person B, person B believes they are talking to person C, and person C believes they are talking to person A.

Furthermore, the model could potentially facilitate communication across time, such as enabling conversations with one's past self. Expanding the structure of the Intersubjective Model in these ways could unlock new possibilities for AIMC.

\subsection{Exploring Designs Tailored to Specific Needs and Conducting Design Research}
Designing AIMC systems based on the Intersubjective Model requires tuning the agents to meet specific user needs. The process of determining this tuning is crucial and necessitates exploration through design research methodologies.

For example, consider the following scenarios:

\begin{itemize}
\item \textbf{Facilitating communication between individuals with social anxiety:} Design research could investigate how to adjust the agent's behavior to create a more comfortable and supportive environment for individuals who experience anxiety in social interactions.

\item \textbf{Supporting creative brainstorming sessions:} Agents could be designed to stimulate idea generation and facilitate the exploration of unconventional ideas by selectively filtering or transforming information shared between participants.

\item \textbf{Mediating conflicts in online discussions:} Design research could explore how to train agents to identify and de-escalate conflicts by promoting empathy and understanding between participants with opposing viewpoints.

\end{itemize}
Through such design research efforts, we can establish design guidelines for AIMC systems that cater to diverse communication needs and contexts.

\subsection{Towards Intersubjective Reality}

In this study, we proposed the Intersubjective Model and conducted practical implementation for text-based communication. However, the potential for this model extends beyond text-based communication to other media forms, particularly in areas such as shared virtual environments and computer-mediated reality~\cite{mann260}. By designing these environments without assuming an objective reality shared by all users, we can remove traditional design constraints and create new types of adaptable, subjective environments. Instead of all users experiencing a single, unified environment, each user could experience a unique environment where their actions and interactions are transmitted across these distinct spaces.

Research in virtual reality has begun exploring this type of intersubjective design on a smaller scale, such as by adjusting for spatial discrepancies~\cite{Yoon2022Dissimilar, Gronbaek2023Partially, Dongsik2015Spacetime}, temporal shifts~\cite{Haijun2018Spacetime, Portia2024Late}, and even avatar-based adaptations~\cite{hasler2017real}. While these studies generally focus on low-magnitude transformations, the design space for higher-magnitude reality asymmetry has yet to be fully explored.

With advancing technologies such as LLMs and video generation, we can begin developing systems where these components function as world models, capable of dynamically generating personalized virtual realities. This would allow for an environment akin to the Intersubjective Model as proposed for text-based communication in this paper—an optimized, customized reality for each individual, where select aspects are extracted and shared. In such a system, reality itself can be individually tailored while enabling shared experiences that promote interpersonal understanding.

Achieving complete subjectivity, where each individual experiences their own fully optimized environment, could risk creating a dystopian world that limits shared experiences and understanding. On the other hand, a fully objective reality would remain unchanged from traditional designs and could perpetuate existing issues arising from a one-size-fits-all environment. Thus, the challenge is to explore designs that balance between objectivity and subjectivity, seeking an intersubjective reality that allows for personalized experiences while maintaining enough shared elements to foster meaningful human connection. This design approach lies between the red and blue pills in The Matrix: neither purely subjective nor entirely objective but a blended reality designed to enhance human experience and interaction.

\section{Conclusion}

This paper introduced the \emph{Intersubjective Model}, a novel framework for AI-mediated communication that shifts away from the traditional assumption of a shared objective environment. By enabling individuals to communicate through AI agents within personalized subjective spaces, the model opens up a vast design space for modulating and enhancing human interaction. We presented the theoretical underpinnings of this model, drawing connections to existing communication theories, and outlined a basic system architecture for its implementation. While this work provides a foundational understanding of the Intersubjective Model, further research is crucial to explore its full potential and address the associated challenges. We believe that investigating the proposed future research directions will lead to a deeper understanding of intersubjectivity in communication and pave the way for innovative AI-mediated communication technologies that enhance human connection in diverse and meaningful ways.

\bibliographystyle{unsrt}  
\bibliography{preprint}

\end{document}